\documentclass[conference]{IEEEtran}
\IEEEoverridecommandlockouts
\usepackage{cite}
\usepackage{amsmath,amssymb,amsfonts}
\usepackage{algorithmic}
\usepackage{dblfloatfix}    
\usepackage{graphicx}
\usepackage{textcomp}
\usepackage{xcolor}
\usepackage{setspace}

\def\BibTeX{{\rm B\kern-.05em{\sc i\kern-.025em b}\kern-.08em
    T\kern-.1667em\lower.7ex\hbox{E}\kern-.125emX}}
\begin{document}

\title{Delay-Doppler Domain Pulse Design for OTFS-NOMA 
}

\setstretch{0.98}

\author{\IEEEauthorblockN{
         Michel Kulhandjian\IEEEauthorrefmark{1}, 
         Hovannes Kulhandjian\IEEEauthorrefmark{2},
         Gunes Karabulut Kurt\IEEEauthorrefmark{3}, 
         Halim Yanikomeroglu\IEEEauthorrefmark{4} %
    } 
     \IEEEauthorblockA{
   \IEEEauthorrefmark{1}Department of Electrical and Computer Engineering, Rice University, Houston, TX, USA \\
    \IEEEauthorrefmark{2}Department of Electrical and Computer Engineering,
			California State University, Fresno, CA, USA\\
    \IEEEauthorrefmark{3}Department of Electrical Engineering,
			Polytechnique Montreal, Montreal, QC, Canada\\
    \IEEEauthorrefmark{4}Systems and Computer Engineering, Carleton University, Ottawa, ON, Canada\\
        \vspace{-20pt} } }


\maketitle

\begin{abstract}
We address the challenge of developing an orthogonal time-frequency space (OTFS)-based non-orthogonal multiple access (NOMA) system where each user is modulated using orthogonal pulses in the delay-Doppler domain. Building upon the concept of the sufficient (bi)orthogonality train-pulse \cite{Lin2022}, we extend this idea by introducing Hermite functions, known for their orthogonality properties. Simulation results demonstrate that our proposed Hermite functions outperform the traditional OTFS-NOMA schemes, including power-domain (PDM) NOMA and code-domain (CDM) NOMA, in terms of bit error rate (BER) over a high-mobility channel. The algorithm's complexity is minimal, primarily involving the demodulation of OTFS. The spectrum efficiency of Hermite-based OTFS-NOMA is $K$ times that of OTFS-CDM-NOMA scheme, where $K$ is the spreading length of the NOMA waveform.
\end{abstract}


\section{Introduction}

6G and beyond networks will be designed to meet a wide range of demands from an increasingly connected society in a sustainable manner \cite{ITU}. These networks need to support a diverse mix of applications, from high-speed mobile internet access through satellites and IoT connectivity to mission-critical communications and virtual reality. Their requirements networks are multifaceted, reflecting the need for significant improvements in capacity, performance, and spectral efficiency \cite{Le2021}, to serve in the presence of high mobility, multipath propagation, and varying channel conditions.



Orthogonal time-frequency space (OTFS), proposed in \cite{7925924},  is a promising approach to provide the fundamental basis of increased spectral efficiency \cite{Lin2022}. It is a two-dimensional signal processing framework designed to address challenges associated with time-varying wireless communication channels. In traditional wireless communication systems, signals are often distorted due to channel variations over time, resulting in signal fading and degradation. OTFS aims to mitigate these effects by representing signals in a transformed domain that simultaneously captures both time and frequency variations by carrying the information in the delay-Doppler domain. Hence, in OTFS, the delay-Doppler domain is employed as the basis for signal representation. This domain accounts for variations in both time delay and Doppler frequency shift, providing a comprehensive view of the channel dynamics. By transforming signals into the delay-Doppler domain, OTFS enables the separation of multipath components and while offering improved resilience against the time-varying nature of wireless channels. 

The key advantage of OTFS lies in its ability to handle challenging channel conditions, such as high-mobility scenarios, where traditional approaches may struggle.

OTFS can provide enhanced performance in terms of signal recovery and communication reliability, particularly when the channel experiences rapid changes over time. The utilization of OTFS has been explored in various communication scenarios, including multiple-input multiple-output (MIMO) systems and non-orthogonal multiple access (NOMA) schemes. Recent interest in applying OTFS principles to satellite channels, particularly in low earth orbit (LEO) satellite communication, has been discussed in \cite{Shi2023}. 

NOMA is a multiple-access technique employed in wireless communication systems to enhance spectral efficiency by allowing multiple users to share the same time-frequency resource. In traditional orthogonal multiple access schemes, such as orthogonal frequency division multiple access (OFDMA), different users are assigned orthogonal resources to avoid interference. NOMA, however, challenges this orthogonality principle and allows multiple users to share the same resources non-orthogonally. The key idea behind NOMA is to utilize the power domain for distinguishing users, allocating different power levels to users accessing the same time-frequency resource. Users are distinguished by assigning them different power levels rather than different time or frequency slots. This non-orthogonal resource allocation enables simultaneous transmissions from multiple users in the same time-frequency resource, improving spectral efficiency. 

In NOMA, users with better channel conditions are allocated higher power levels, allowing them to decode their signals with less interference. Users with poorer channel conditions receive lower power levels and experience interference from the stronger signals. Advanced signal processing techniques, such as successive interference cancellation (SIC) or joint decoding, are employed at the receiver to recover the signals of users with lower power levels. NOMA has gained attention as a promising technique to address the increasing demand for efficient use of limited wireless spectrum resources. It is particularly relevant in scenarios with varying channel conditions and diverse user requirements. NOMA is being explored in various communication technologies, including 5G and beyond, to achieve higher throughput, lower latency, and improved connectivity in wireless networks.

\begin{figure}[tb]
\centering
\begin{center}
\hspace*{-.5cm}		\hspace{-0.3cm}\includegraphics[width=3.0 in]{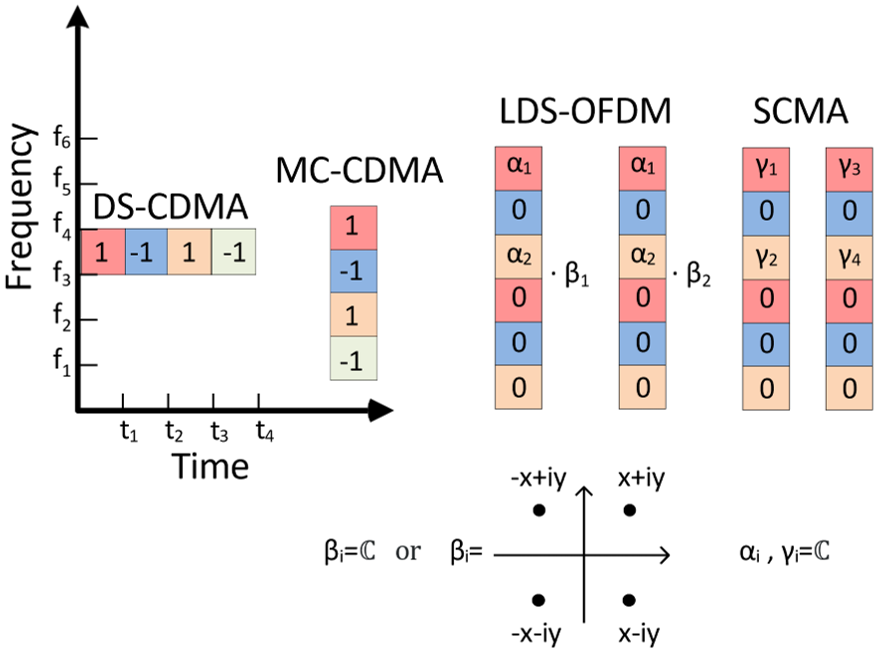}\vspace{-3mm}
\caption{A comparison of multiple access techniques.} \label{comparison02}
\end{center}
\vspace{-0.3cm}
\end{figure}

The sparse spreading non-orthogonal multiple access (NOMA) can be classified into power-domain NOMA (PDM-NOMA) \cite{MichelHanzo2020, Michel2021} and code-domain NOMA (CDM-NOMA) \cite{Dai2015}. Some notable instances of CDM-NOMA include low-density spreading aided CDMA (LDS-CDMA) \cite{Michel2022}, low-density spreading assisted orthogonal frequency-division multiplexing (LDS-OFDM) \cite{Razavi2012}, sparse code multiple access (SCMA) \cite{Nikopour2013, michelPIMRC2017}, and multi-user shared access (MUSA) \cite{Yuan2016}. Among the variations of NOMA, LDS can be considered a special case of SCMA, characterized by sparse codebooks. Each codebook is expressed as the Kronecker product of a sparse sequence denoted by $\mathbf{s}_j$ and a constellation set of order $M$. Specifically, we have
\begin{equation}
\mathbf{X}_j = [\mathbf{s}_j \beta_1, \mathbf{s}_j \beta_2, \dots, \mathbf{s}_j \beta_M],
\end{equation}
where ${\beta_1, \beta_2, \dots, \beta_M }$ indicates a constellation set. Consequently, the rank of the users' LDS codebooks, $\mathbf{X}_j$, is equal to one. However, this is not the case for the users' SCMA codebooks. The rank of SCMA codebooks is higher than one and is equal to the number of non-zero values in the SCMA waveforms. The comparison between direct sequence CDMA (DS-CDMA), multicarrier CDMA (MC-CDMA), LDS-OFDM, and SCMA is illustrated in Fig. \ref{comparison02}.

Research in the exploration of the synergy between NOMA and OTFS is ongoing, with efforts focused on optimizing these techniques to address practical implementation challenges, such as computational complexity and real-time processing requirements. So far, mostly the power domain use of NOMA has been considered \cite{8786203, 10218331}. Code domain extensions are also proposed \cite{Deka2021, 10022044}, but the waveform type is kept the same in the existing approaches. 
\begin{figure*}[!t]
    \centering
    \includegraphics[width=16.0cm]{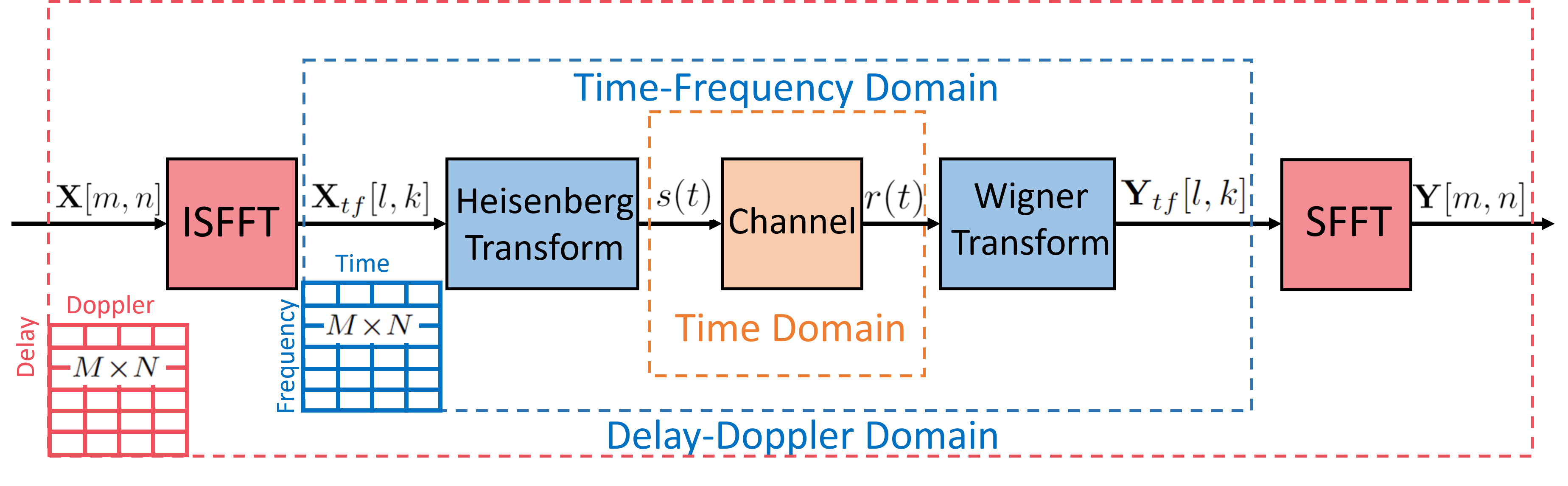}\vspace{-4mm}
    \caption{OTFS block system.}
    \label{OTFS}    
\end{figure*}
The primary contribution of our paper lies in the development of an OTFS-based NOMA scheme where individual users are assigned distinct orthogonal pulses, enabling them to share the same bin in the delay-Doppler grid. This achievement is facilitated by leveraging the well-established Hermite functions known for their orthogonal properties. It is crucial to highlight that our proposed scheme differs from traditional NOMA approaches that involve designing novel NOMA codebooks for user superimposition using different codes. Instead, our emphasis is on the development of orthogonal pulses. Our modulation scheme adopts the delay-Doppler pulse-train design \cite{Lin2022}, opting for Hermite functions over square-root Nyquist pulses. Our proposed schemes surpassed both OTFS-PDM-NOMA and OTFS-CDM-NOMA in terms of bit error rate (BER) and demonstrated improvements in both complexity and spectral efficiency.   

The rest of this paper is organized as follows. in the following section, we provide the preliminary information to present the framework. The proposed delay-Doppler post design is presented in Section \ref{PulseDesign}. The numerical results, illustrating the complexity and spectral efficiency improvements resulting from the utilization of Hermite functions due to their orthogonality property, are presented in Section \ref{Simulation}. The conclusions are drawn in Section \ref{Conclusion}.



\section{Preliminaries}
\label{Preliminaries}
\subsection{OTFS}
In this section, we provide a brief overview of the delay-Doppler (DD) representation, modulation, and demodulation processes within the OTFS system. We consider a discrete-time baseband model where a continuous-time OTFS signal is sampled at a sampling frequency $f_s = \frac{1}{T_s} = B$, with $T_s$ denoting the sampling
interval and $B$ representing the bandwidth. Each OTFS frame comprises of $NM$ samples organized into $N$ blocks, each containing $M$ samples. Consequently, the duration of the resulting OTFS frame can be calculated as $T_f = N M T_s = NT$, where $T = M T_s$ signifies the duration of each block. At the transmitter, the $NM$ information symbols, drawn from a modulation alphabet (e.g., QAM, etc.), are arranged in a delay-Doppler grid as follows
\begin{equation}
\Gamma = \left \{ \left (m\Delta \tau, n\Delta \nu \right),  m = 0,\dots, M-1; n = 0, \dots, N-1  \right \},
\end{equation}
\noindent where $M$ and $N$ represent the number of delays and Doppler bins, $\Delta \tau = \frac{1}{M \Delta f}$ and $\Delta \nu = \frac{1}{N T}$ serve as the delay and Doppler resolution of the delay-Doppler grid,  $\Delta f = 1/T$, respectively. The bandwidth $B$ and the frame duration $T_f$ in terms of delay and Doppler resolution are given as 
\begin{equation}
B = \frac{1}{\Delta \tau} = M \Delta f, \: \: T_f = \frac{1}{\Delta \nu} = NT.
\end{equation}

In OTFS, the DD domain grid $\Gamma$ undergoes a conversion into the time-frequency (TF) domain grid $\Lambda$ through the inverse symplectic fast Fourier transform (ISFFT), given by
\begin{equation}
\Lambda = \left \{ \left (l\Delta f, k T \right),  l = 0,\dots, M-1; k = 0, \dots, N-1  \right \}
\end{equation}
\noindent where the symbol interval $T$ (in seconds) and subcarrier interval $\Delta f$ (Hz) serve as the time and frequency resolutions of the TF grid $\Lambda$, respectively.

 The selection of parameters $N$ and $M$ depends on the delay and Doppler conditions of the underlying channel for the transmission of $NM$ symbols over a frame of duration $T_f$ and bandwidth $B$. The design parameters are chosen such that $\tau_{max} < \frac{1}{\Delta f}$, and $\nu_{max} < \frac{1}{T}$,  where $\tau_{max}$ and $\nu_{max}$ represent the delay and Doppler spread, respectively. For example, a channel with a high Doppler spread $\nu_{max}$ would necessitate a higher $\frac{1}{T} = N \Delta \nu$, leading to smaller $T$ and larger $\Delta f$, implying larger $N$ and smaller $M$. Similarly, if the channel exhibits a higher delay spread $\tau_{max}$, it would require a higher $\frac{1}{\Delta f} = M \Delta \tau$, resulting in larger $T$ and smaller $\frac{1}{\Delta f}$, suggesting smaller $N$ and larger $M$. Another design constraint for OTFS systems is 
 stems from the assumption that the multipath channel parameters remain constant throughout the frame duration, denoted as
 $T_f$ (e.g., $10 - 20$ ms).

At the transmitter, the elements of $\Gamma$ are transformed into a DD domain matrix $\mathbf{X} \in \mathbb{C}^{M \times N}$ with entries $\mathbf{X}[m,n]$, where $m = 0, \dots, M-1$, and $n = 0,\dots, N-1$. As illustrated in Fig. \ref{OTFS}, the transmitter initially maps the symbols $\mathbf{X}[m,n]$ to be transmitted into $NM$ samples on the TF grid using the ISFFT, given by
\begin{equation}
\mathbf{X}_{tf}[l,k] = \frac{1}{\sqrt{N M}} \sum_{n = 0}^{N-1}\sum_{m = 0}^{M-1} \mathbf{X}[m,n] e^{j2\pi (\frac{nk}{N}- \frac{ml}{M})} ,
\label{ISFFT}
\end{equation}
for $l = 0, \dots, M-1$, $k=0, \dots, N-1$, where $\mathbf{X}_{tf}[l,k] \in \mathbb{C}^{M \times N}$ represents the TF domain transmitted matrix.
Note that the distinction between OFDM and OTFS transmission lies in the fact that OTFS is a 2-D modulation technique, whereas OFDM is a 1-D modulation scheme. Following the 2-D modulation in the time-frequency domain represented by $\mathbf{X}_{tf}[l,k]$, the signal needs to be converted to the continuous-time waveform $s(t)$. This transformation is accomplished at the transmitter using the Heisenberg transform and a basis pulse denoted by $g_{tx}(t)$, expressed as
\begin{equation}
s(t)  =  \sum_{k = 0}^{N-1}\sum_{l = 0}^{M-1} \mathbf{X}_{tf}[l,k] g_{tx}(t-kT) e^{j2\pi l \Delta f (t-kT)}.
\label{Heisenberg}
\end{equation}
In the scenario of a high-mobility channel, the signal $s(t)$  undergoes the effects of a time-varying channel represented by the delay-Doppler response $h(\tau, \nu)$, as follows
\begin{equation}
r(t)  =  \int \int h(\tau, \nu) s(t-\tau) e^{j2\pi \nu  (t-\tau)} d\tau d\nu + n(t),
\end{equation}
\noindent where $n(t)$ is an additive white Gaussian noise (AWGN). 
At the receiver, there exists an inverse transform that maps the time-domain received signal $r(t)$ back to the time-frequency domain $\mathbf{Y}_{tf}[n,m]$ using an appropriate basis pulse $g_{rx}(t)$. In Fig. \ref{OTFS}, the signal $r(t)$ undergoes initial processing through a matched filter, which computes the cross-ambiguity function $\mathcal{A}_{g_{rx}, r}(f,t)$ as follows
\begin{equation}
Y(f,t) = \mathcal{A}_{g_{rx}, r}(f,t) =  \int^{\infty}_{\infty} r(t') g^*_{rx}(t'-t) e^{-j2\pi f  (t'-t)} dt' ,
\label{crossAmb}
\end{equation}
\noindent where $^*$ indicates the complex operation. Following the sampling of the matched filter output at regular intervals of $(l\Delta f, kT)$, a 2-D time-frequency received samples matrix $\mathbf{Y}_{tf} \in \mathbb{C}^{M \times N}$ is obtained, with entries given by 
\begin{equation}
\mathbf{Y}_{tf}[l,k] = Y(f,t)|_{f = l\Delta f, t = kT},
\end{equation}
\noindent for $l=0, \dots, M-1$, and $k=0,\dots, N-1$. This inverse transform is commonly known as the Wigner transform. Finally, the symplectic fast Fourier transform (SFFT) is applied on $\mathbf{Y}_{tf}[l,k]$ to obtain the DD received matrix $\mathbf{Y}\in \mathbb{C}^{M \times N}$ with entries as follows
\begin{equation}
\mathbf{Y}[m,n] = \frac{1}{\sqrt{N M}} \sum_{k = 0}^{N-1}\sum_{l = 0}^{M-1} \mathbf{Y}_{tf}[l,k] e^{-j2\pi (\frac{nk}{N}- \frac{ml}{M})} .
\end{equation}
For more in-depth information on OTFS, consult existing publications, such as those referenced in [6] and [13]. 

In OTFS, a fundamental challenge in designing a multicarrier modulation scheme involves determining the pulses $g_{tx}(t)$ and $g_{rx}(t)$ that satisfy the (bi)orthogonal property concerning the symbol interval $T$ and subcarrier interval $\Delta f$ in the TF domain. Let the joint time-frequency resolution be denoted as $\mathcal{R}$. According to Weyl-Heisenberg (WH) or Gabor frame theory, sets of (bi)orthogonal WH functions only exist for $\mathcal{R} \geq 1$ \cite{Lin2022}, and as a result, most multicarrier modulation schemes are designed with $\mathcal{R} \geq 1$. 

In accordance with WH frame theory, the presence of (bi)orthogonal WH pulses is contingent on $\mathcal{R}$ as outlined below
\begin{itemize}
\item $\mathcal{R} = 1$ (critical sampling): Orthogonal WH pulses exist, featuring infinite energy spread either in the time or frequency domain as per the Balian-Low theory. Consequently, these pulses lack well-localized characteristics in the TF domain.

\item $\mathcal{R} > 1$ (under-critical sampling): Well-localized orthogonal or biorthogonal WH pulses exist if $\mathcal{R}$ significantly exceeds 1.

\item $\mathcal{R} < 1$ (over-critical sampling): Neither orthogonal nor biorthogonal WH pulses are present.
\end{itemize}
Observing the TF domain resolution $\mathcal{R}_{TF} = \Delta f T = 1$, it becomes evident that any pulse satisfying WH theory possesses infinite pulse energy. Furthermore, in the DD domain, where $\mathcal{R}_{DD} = \frac{1}{NT} < 1$, (bi)orthogonal WH pulses, according to WH frame theory, do not exist either. 
However, by relaxing the global (bi)orthogonality condition and narrowing our focus to a limited set of symbols and subcarriers, this approach facilitates pulse design capable of meeting the orthogonality requirement.
As per \cite{Lin2022}, such DD pulses that fulfill the adequate (bi)orthogonality condition exist. If we consider $u(t)$ as a pulse-train given by
\begin{equation}
u(t) = \sum_{n=0}^{N-1} a(t-nT),
\end{equation}
\noindent where the subpulse $a(t)$ is a square-root Nyquist pulse parameterized by its zero-inter-symbol interference (ISI) delay $T/M$ bins in DD plane, it has been demonstrated that $u(t)$ satisfies the sufficient orthogonality property of the cross-ambiguity function as defined in (\ref{crossAmb}). This can be expressed as \cite{Lin2022}
\begin{equation}
\mathcal{A}_{u,u}\left (m \frac{T}{M} , n\frac{1}{NT}\right ) = \delta(m) \delta(n),
\label{DDambiguity}
\end{equation}
for $|m| \leq M -1$ and $|n| \leq N -1$.

 \subsection{OTFS-NOMA}
  The OTFS framework has found application in various schemes, including NOMA. In the PDM-NOMA scenario, stationary users employ the traditional NOMA scheme, while high-mobility users are modulated using the OTFS scheme. For the mobile users, they are positioned in the DD grid, where the ISFFT is initially applied, followed by the Heisenberg transformation \cite{Deng2021}.
In the context of OTFS-CDM-NOMA, the spreading codes of users are superimposed and placed in the DD grid, spanning the length of the spreading code. Following this, the ISFFT and Heisenberg transformations are applied before transmission. At the receiver, the initial step involves OTFS demodulation, followed by NOMA decoding \cite{Deka2021}. 
As an illustration, let's consider a downlink scenario involving $J$ receiving users and one transmitting base station (BS). Initially, the BS superimposes the signals from the $J$ users as follows
\begin{equation}
\mathbf{s}' = \sum_{j = 1}^J \mathbf{s}_j,
\end{equation}
\noindent where the summation is performed bin-wise. Each bin of the superimposed signal $\mathbf{s}'$ is then placed in the DD grid. Subsequently, it undergoes OTFS modulation, encompassing the processes described in (\ref{ISFFT}) - (\ref{Heisenberg}).

\section{Delay-Doppler Pulse Design}
\label{PulseDesign}
\begin{figure}[h]
    \centering
    \includegraphics[width=8.9cm]{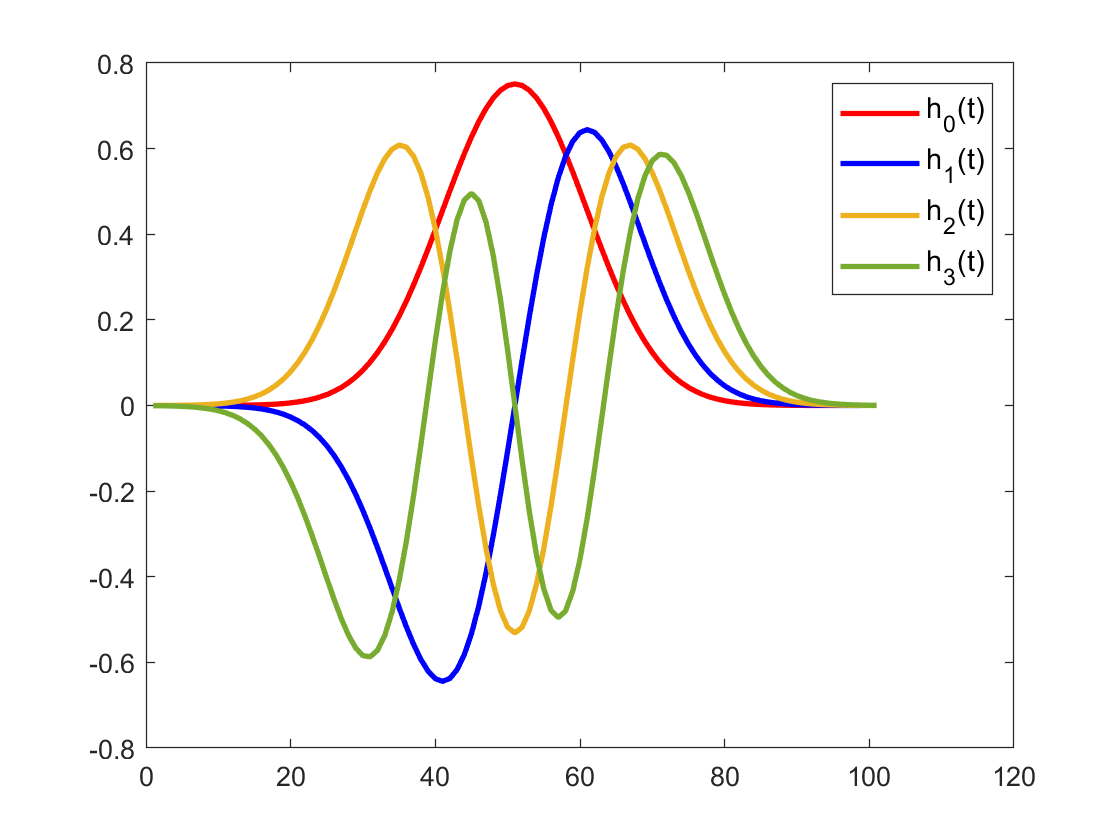}\vspace{-5mm}
    \caption{Hermite Functions of order $k=0,1,2,3$.}
    \label{Hermites}    
\end{figure}

In this section, we explore a novel OTFS-based NOMA scheme where multiple users are superimposed not at the level of spreading codes (e.g., LDS, SCMA) but at the pulse $g_{tx}(t)$ level. This approach leverages an additional diversity scheme. One potential solution entails designing pulses $g_{tx}(t)$ for each user in a way that ensures they exhibit orthogonality properties. To achieve this desired orthogonality property, Hermite functions are employed, given that they form an orthogonal basis for $L_2(\mathbb{R}^2)$. Moreover, in addition to their orthogonality property, Hermite functions also demonstrate an excellent time-frequency localization property. The construction of Hermite functions involves a Hermite polynomial, a Gaussian window, and a normalization. A Hermite function of order $k$ is defined as 
\begin{equation}
h_k(t) \triangleq \frac{H_k(t)e^{-\frac{t^2}{2}}}{\sqrt{2^k k!\sqrt{\pi}}},
\label{Hermite}
\end{equation}
\noindent where $H_k$ is the Hermite polynomial of the $k$-th degree given by
\begin{equation}
H_k(t) = (-1)^k e^{t^2} \frac{d^k}{dt^k}(e^{-t^2}).
\end{equation}
The Hermite functions in (\ref{Hermite}) are orthogonal
\begin{equation}
| \langle h_i(t), h_j(t) \rangle | = \left | \int h_i(t) h_j(t) dt \right | = \delta_{i, j},
\end{equation}
\noindent where $\delta_{i, j}$ denotes the Kronecker delta function. Thus, the superposition of NOMA users on the DD grid is orthogonal because each user is associated with a different Hermite order function. The first four orders of Hermite functions are illustrated in Fig. \ref{Hermites}.
\vspace{-0.6cm}

\begin{figure}[h]
    \centering
    \includegraphics[width=8.5cm]{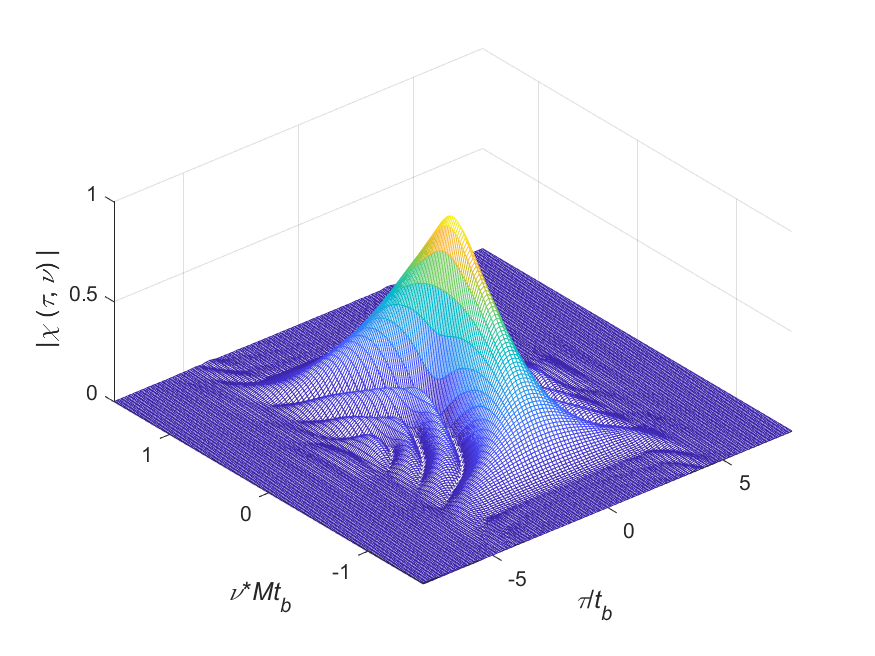}
    \vspace{-6mm}
    \caption{RRC pulse.}
    \label{RRC}    
\end{figure}

\begin{figure}[h]
    \centering
    \includegraphics[width=8.5cm]{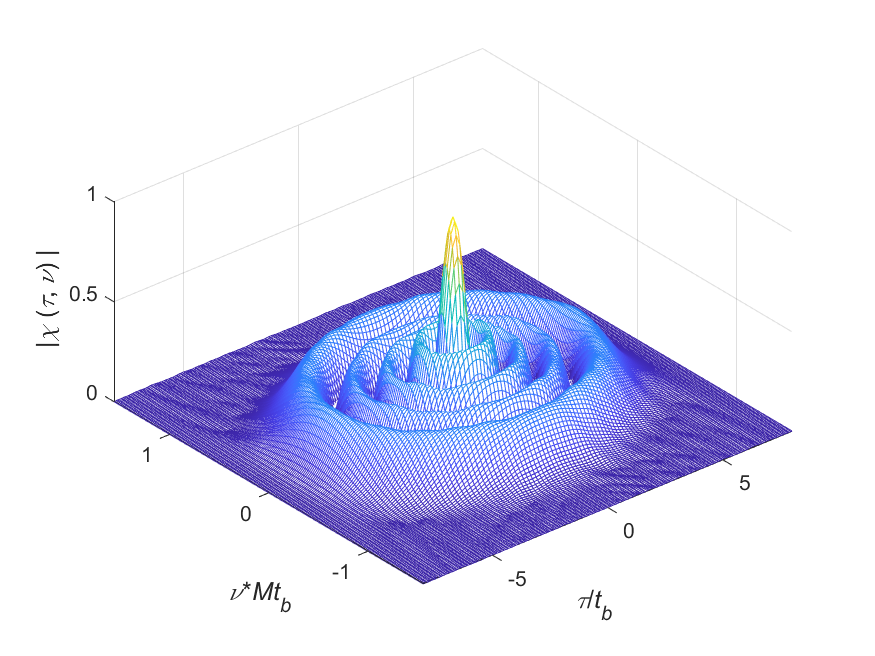}\vspace{-6mm}
    \caption{ Fourth-order Hermite pulse.}
    \label{Hermite4}    
\end{figure}

\begin{figure}[h]
    \centering
    \includegraphics[width=8.5cm]{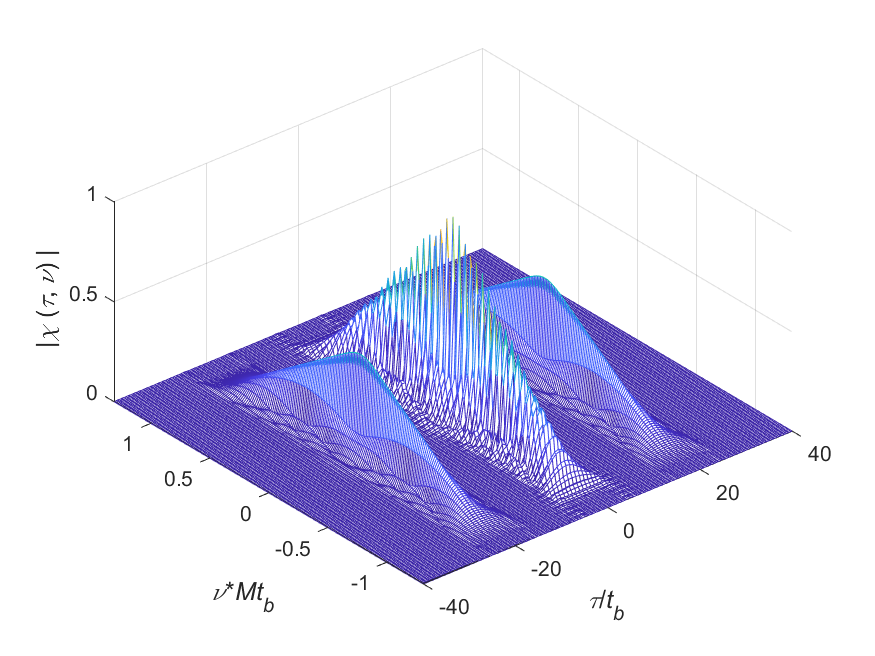}\vspace{-6mm}
    \caption{RRC two pulses.}
    \label{RRC2Pulses}    
\end{figure}

\begin{figure}[h]
    \centering
    \includegraphics[width=8.5cm]{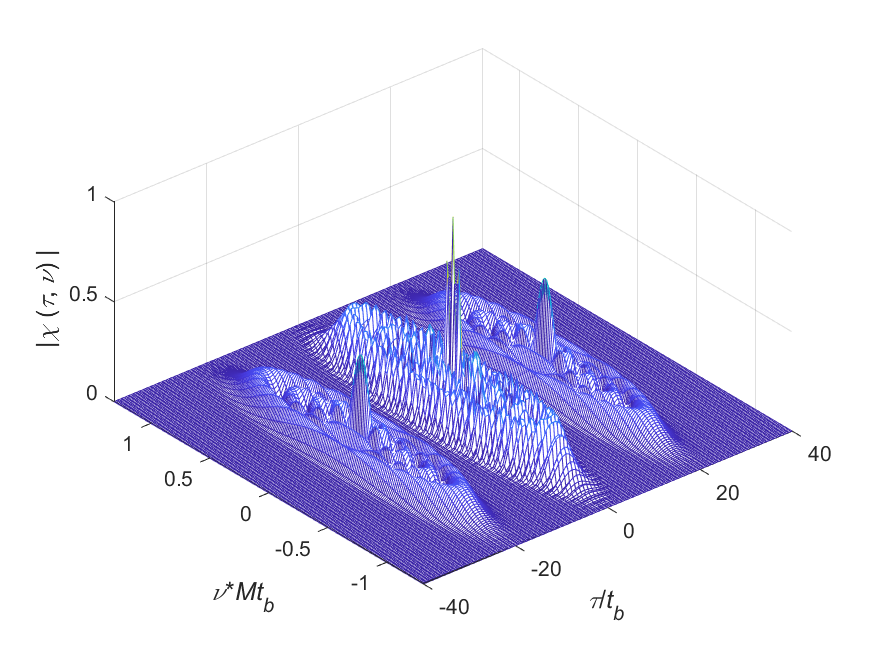}\vspace{-6mm}
    \caption{ Fourth-order Hermite two pulses.}
    \label{Hermite4Twopulses}    
\end{figure}

In our pulse design procedure, we will adapt the DD domain pulse-train $u(t)$ that satisfies the sufficient (bi)orthogonality condition (\ref{DDambiguity}) as outlined in \cite{Lin2022}. The distinction in our approach lies in the use of Hermite functions instead of zero-ISI square-root Nyquist pulses (e.g., root-raised cosine (RRC), etc.), as applied in \cite{Lin2022}. The possibility of superposing different users on the DD grid arises from the orthogonality of Hermite functions of different orders assigned to distinct users. However, it is essential to highlight that Hermite functions, unlike RRC pulses, do not achieve zero and cannot be staggered, as explained in \cite{Lin2022}. To address this, we define the Hermite functions as follows
\begin{equation}
h_k(t) = 
\begin{cases}
  h_k(t) & \text{$0\leq t \leq \frac{T}{M}$} \\
  0 & \text{otherwise}
\end{cases}.
\label{HermiteISI}
\end{equation}
By doing so, (\ref{HermiteISI}) ensures that Hermite functions satisfy the sufficient (bi)orthogonality condition in (\ref{DDambiguity}). A three-dimensional plot of the ambiguity functions in (\ref{DDambiguity}) is depicted in Figs. \ref{RRC} - \ref{Hermite4Twopulses} for RRC and Hermite pulses of order 4.

\vspace{-0mm}

\vspace{-0mm}

 
\section{Simulation}
\label{Simulation}
\begin{figure}[h]
    \centering
    \includegraphics[width=8.9cm]{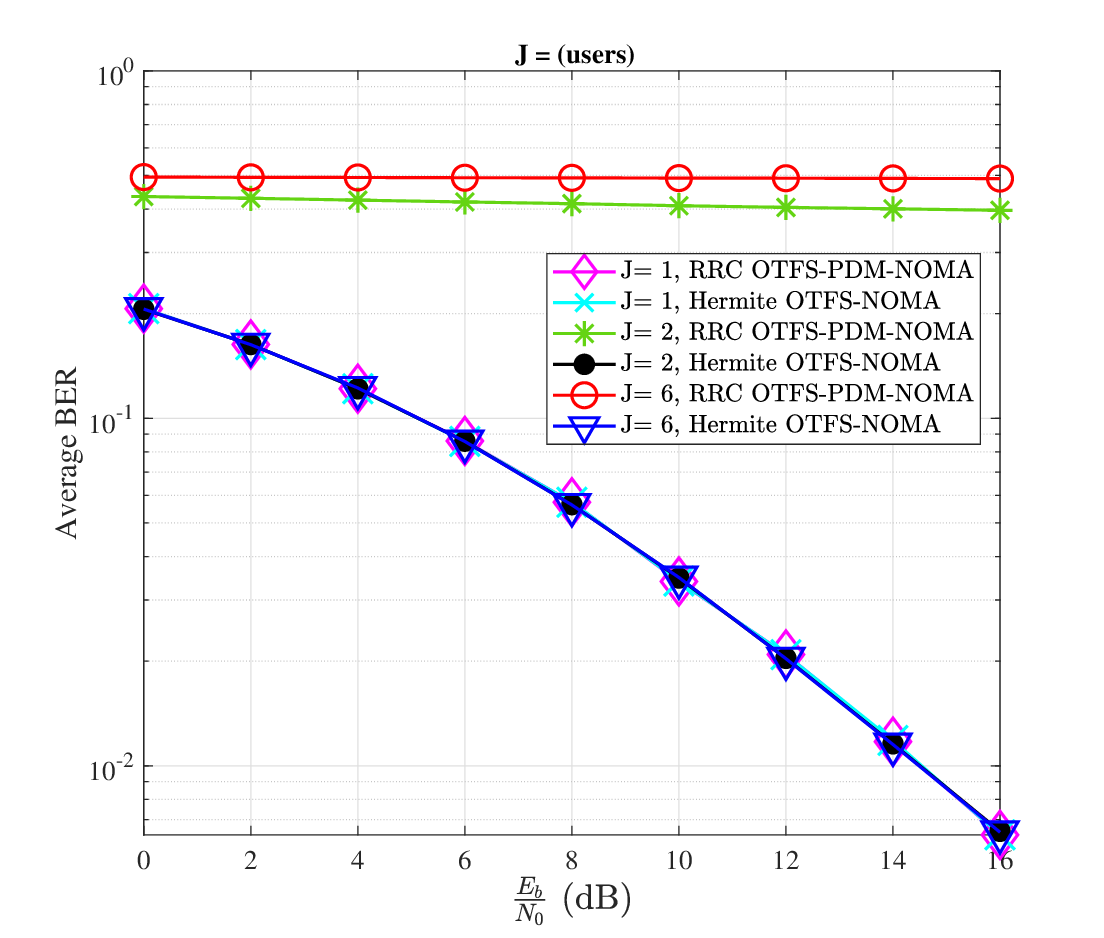}\vspace{-3mm}
    \caption{BER performance of Hermite-OTFS-NOMA versus RRC-OTFS-PDM-NOMA.}
    \label{1User}    
\end{figure}

In this section, we assess the performance of our proposed Hermite-based OTFS-NOMA scheme compared to the RRC-based OTFS-NOMA. Our simulations focus on the uplink scenario with a multiuser case received at the BS. We incorporate both OTFS-PDM-NOMA and OTFS-CDM-NOMA, specifically using SCMA. For the RRC-based scheme, multiuser detection is carried out using SIC estimator and the message passing algorithm (MPA). In contrast, for the Hermite-based scheme, user detection is accomplished through matched filtering, leveraging the orthogonality of users at the DD grid level. The DD grid is constructed with $M=4$, $N = 2$, and a $4$-QAM constellation for transmitted symbols. In the simulations, the maximum speed is $500$ km/h, and the roll-off factor of the RRC pulse is chosen to be $0.2$. For the SCMA scenario we adopted OTFS-SCMA scheme outlined in \cite{Deka2021}.
Figure \ref{1User} illustrates the BER performance comparison between the Hermite-based OTFS-NOMA and RRC-based OTFS-PDM-NOMA schemes. In the single-user scenario, the performance is comparable; however, OTFS-PDM-NOMA degrades with the inclusion of more users.

\begin{figure}[h]
    \centering
    \includegraphics[width=8.9cm]{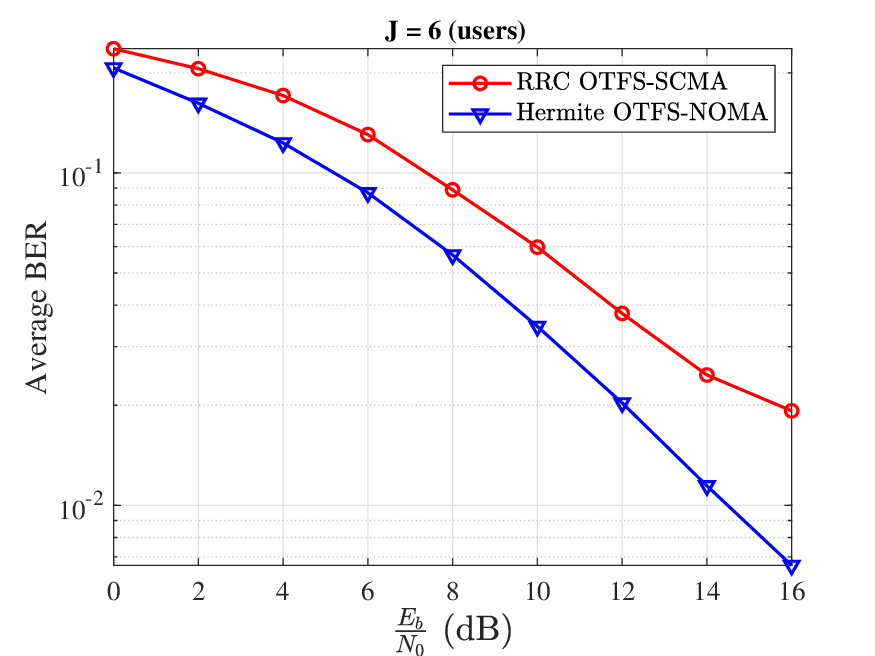}\vspace{-3mm}
    \caption{BER performance of Hermite-OTFS-NOMA versus RRC-OTFS-SCMA.}
    \label{SCMA01}    
\end{figure}

Figure \ref{SCMA01} portrays that OTFS-SCMA outperforms OTFS-PDM-NOMA, but it falls short of surpassing the Hermite-based OTFS-NOMA. The MPA introduces additional processing complexity of $\mathcal{O}(\mathcal{M}^{d_f})$. This complexity grows exponentially with both the size of the symbol alphabet $\mathcal{M}$ and the number of non-zero positions in the spreading waveform $d_f$. In terms of spectrum efficiency, both OTFS-PDM-NOMA and Hermite-OTFS-NOMA demonstrate similar efficiency, but OTFS-PDM-NOMA underperforms in multiuser scenarios. The spectrum efficiency of Hermite-OTFS-NOMA is $K$ times that of OTFS-SCMA, where $K$ is the spreading length of the SCMA waveform. The Hermite-based scheme exhibits superior performance compared to the RRC-based OTFS-NOMA, encompassing both OTFS-PDM-NOMA and OTFS-CDM-NOMA schemes. The utilization of Hermite functions enhances the orthogonality among different users, introducing additional diversity and thereby improving user separation.

\section{Conclusion}
\label{Conclusion}
In conclusion, this paper addresses the complex challenge of developing an orthogonal time-frequency space (OTFS)-based non-orthogonal multiple access (NOMA) system. In our approach, each user is modulated using orthogonal pulses in the delay-Doppler domain. We extend the existing concept of the sufficient (bi)orthogonality train-pulse by introducing Hermite functions, leveraging their orthogonality properties.

Simulation results affirm the superiority of our proposed Hermite functions over the traditional OTFS-NOMA scheme, particularly evident in the realm of bit error rate (BER) over high-mobility channels. Despite the advanced capabilities demonstrated, the algorithm maintains minimal complexity, primarily relying on OTFS demodulation. This work contributes valuable insights to the field, showcasing the potential of Hermite functions in optimizing the performance of OTFS-NOMA systems under challenging channel conditions.

\bibliographystyle{IEEEtran}
\bibliography{Ref}

\end{document}